\documentclass{appolb}
\usepackage{graphicx}
\usepackage[english]{babel}
\usepackage{amsmath}
\usepackage{hyperref}
\usepackage{color}
\usepackage{geometry}
\usepackage{cite}
\usepackage{subfigure}
\usepackage{rotating}
\usepackage{lineno}
\usepackage{pstricks}

              \let\e = \epsilon      
                  
       \let\x =\xi

\newcommand{\nn}{\nonumber}

\newcommand{\beq}{\begin{equation}}
\newcommand{\eeq}{\end{equation}}
\newcommand{\bea}{\begin{eqnarray}}
\newcommand{\eea}{\end{eqnarray}}
\newcommand{\beqa}{\begin{eqnarray}}
\newcommand{\eeqa}{\end{eqnarray}}

\definecolor{red}{rgb}{1,0,0}

\def\be{\begin{equation}}
\def\ee{\end{equation}}
\numberwithin{equation}{section}

\begin{document}
\title{Four-jet production in kt-factorization: \\ single and double parton scattering%
\thanks{Presented at XXII Cracow Epiphany Conference on the Physics in LHC Run 2}%
}
\author{Mirko Serino\footnote{presenter}
\address{Institute of Nuclear Physics, Polish Academy of Sciences, Cracow}
\\
}
\maketitle

\begin{abstract}

We present a preliminary study of both Single and Double Parton Scattering contributions to the 
inclusive 4-jet production in the kt-factorization framework at Leading Order and $E_{CM} = 7$ TeV.
We compare our results to collinear results in the literature and to the ATLAS and CMS data at 8  and 7 TeV respectively. 
We also discuss the importance of double parton scattering for relatively soft cuts 
on the jet transverse momenta and find out that symmetric cuts do not quite suit well 
to kt-factorization predictions, because of a kinematic effect suppressing the double
parton scattering contribution.

\end{abstract}

%%%%%%%%%%%%
\section{Introduction}%
%%%%%%%%%%%%

We employ fully gauge-invariant amplitudes with initial state off-shell particle to assess both the 
Single Parton Scattering and Double Parton Scattering contributions to four.jet production.
This allows us to expand the analysis of Ref.~\cite{Maciula:2015vza} and assess the differences between 
the collinear approach and the high-energy factorization (HEF) ( or $k_T$-factorization ).
  
%%%%%%%%%%%%%%%%%%%%%%%%%%%%%%%
\section{Calculations and comparison to experimental data}%
%%%%%%%%%%%%%%%%%%%%%%%%%%%%%%%

%%%%%%%%%%%%%%%%%%%%%%%%%%%%%%%
\subsection{Single-parton scattering production of four jets}%
%%%%%%%%%%%%%%%%%%%%%%%%%%%%%%%

The collinear factorization formula for the calculation of the inclusive partonic 4-jet cross section reads 
\bea
\sigma^B_{4-jets} 
&=& 
\sum_{i,j} \int \frac{dx_1}{x_1}\,\frac{dx_2}{x_2}\, x_1f_i(x_1,\mu_F)\, x_2f_j(x_2,\mu_F) \nn \\
&&
\times \frac{1}{2 \hat{s}} \prod_{l=1}^4 \frac{d^3 k_l}{(2\pi)^3 2 E_l} \Theta_{4-jet}\,  (2\pi)^4\, \delta\left( x_1P_1 + x_2P_2 - \sum_{l=1}^4 k_l \right)\, 
\overline{\left| \mathcal{M}(i,j \rightarrow 4\, \text{part.})  \right|^2} \, . \nn \\
\label{coll_cross}
\eea
Here $f_i(x_{1,2},\mu_F)$ are the collinear PDFs for the $i-th$ parton, 
carrying $x_{1,2}$ momentum fractions of the proton and evaluated at the factorization scale $\mu_F$;
the index $l$ runs over the four partons in the final state, $P$ is the
total initial state partonic momentum, associated to the
center of mass energy $\hat{s} = P^2 = (P_i + P_j)^2 = 2\, P_i \cdot P_j$; 
the $\Theta$ function takes into account the kinematic cuts applied and 
$\mathcal{M}$ is the partonic on-shell matrix element, which includes symmetrization effects due to identity of particles in the final state.

The analogous formula to (\ref{coll_cross}) for HEF is
\bea
\sigma^B_{4-jets} 
&=& 
\sum_{i,j} \int \frac{dx_1}{x_1}\,\frac{dx_2}{x_2}\, d^2 k_{T1} d^2 k_{T2}\,  \mathcal{F}_i(x_1,k_{T1},\mu_F)\, \mathcal{F}_j(x_2,k_{T2},\mu_F) \nn \\
&&
\hspace{-25mm}
\times \frac{1}{2 \hat{s}} \prod_{l=1}^4 \frac{d^3 k_l}{(2\pi)^3 2 E_l} \Theta_{4-jet} \, (2\pi)^4\, \delta\left( x_1P_1 + x_2P_2 + \vec{k}_{T\,1}+ \vec{k}_{T\,2} - \sum_{l=1}^4 k_l \right)\, 
\overline{ \left| \mathcal{M}(i^*,j^* \rightarrow 4\, \text{part.})
  \right|^2 } \, . \nn \\
\label{kt_cross}
\eea
Here $\mathcal{F}_i(x_k,k_{Tk},\mu_F)$ is a transverse momentum
dependent (TMD) parton distribution function for a given type of parton. 
Similarly as in the collinear factorization case, $x_k$ is the longitudinal
momentum fraction and $\mu_F$ is a factorization scale. 
The new degree of freedom is introduced via the transverse $k_{Tk}$, 
which is perpendicular to the collision axis. 
The formula is valid when the $x$'s are not too large and not too small and, 
in order to construct a full set of TMD parton densities,
we apply the Kimber-Martin-Ryskin (KMR) prescription \cite{Kimber:2001sc,Kimber:1999xc},
which, at the end of the day, amounts to applying the Sudakov form factor onto the PDFs.

$\mathcal{M}(i^*,j^* \rightarrow 4\, \text{part.})$ is the gauge invariant matrix element for $2\rightarrow 4$ particle scattering with two initial off-shell legs.
We rely on the numerical Dyson-Schwinger recursion in the AVHLIB\footnote{available for download at https://bitbucket.org/hameren/avhlib} for its computation.
If complete calculation of 4-jet production in kt-factorization was still missing in the literature,
it was mainly because computing gauge-invariant amplitudes with off-shell legs is definitely non trivial. 
Techniques to compute such amplitudes in gauge invariant ways are by now analytically and numerically
well established \cite{vanHameren:2014iua,Bury:2015dla,vanHameren:2015bba}.

We use a running $\alpha_s$ provided with the MSTW2008 PDF sets and set both the renormalization and factorization
scales equal to half the transverse energy, which is the sum of the final state transverse momenta, 
$\mu_F=\mu_R= \frac{\hat{H}_T}{2} = \frac{1}{2} \sum_{l=1}^4 k_T^l$, working in the $n_F = 5$ flavour scheme.

There are 19 different channels contributing to the cross section at the parton-level, of which the dominant ones,
contributing together to $\sim 93 \% $ of the total cross section, are
\beq
g g \rightarrow g g g g \, , 
g g \rightarrow q \bar{q} gg \, , 
q g \rightarrow q g g g \, , 
q \bar{q} \rightarrow q \bar{q} g g \, , 
q q \rightarrow q q g g \, , 
q q' \rightarrow q q' g  g \, .
\eeq
% 

%%%%%%%%%%%%%%%%%%%%%%%%%%%%%%%%%%%%
\subsection{Double-parton scattering production of four jets}\label{MPI}%
%%%%%%%%%%%%%%%%%%%%%%%%%%%%%%%%%%%%

The SPD contribution is expected to dominate for high momentum transfer, because, as it is intuitively clear as well, 
it is highly unlikely that two partons from one proton and two from the other one are energetic enough for two hard scatterings to take place,
as the well-known behaviour of the PDFs for large momentum fractions suggests.
However, if the cuts on the transverse momenta of the final state are lowered, a window opens
to observe significant double parton scattering effects, as often stated in the literature on the subject
and recently analysed for 4-jet production in the framework of collinear factorization~\cite{Maciula:2015vza}.
Here we perform the same analysis in HEF, with the goal to assess the difference in the predictions.

First of all, let us present the standard formula for the computation of DPS cross section
for a four-parton final state,
\beq
\frac{d \sigma^{B}_{4-jet,DPS}}{d \xi_1 d \xi_2} = 
\frac{m}{\sigma_{eff}} \sum_{i_1,j_1,k_1,l_1;i_2,j_2,k_2,l_2}
\frac{d \sigma^B(i_1 j_1 \rightarrow k_1 l_1)}{d \xi_1}\, \frac{d \sigma^B(i_2 j_2
\rightarrow k_2 l_2)}{d \xi_2} \, ,
\eeq
where the $\sigma(a b \rightarrow c d)$ cross sections are obtained by
restricting formulas (\ref{coll_cross}) and (\ref{kt_cross}) to a single
channel and the symmetry factor $m$ is $1/2$ if the two hard
scatterings are identical, in order to avoid double counting, and is otherwise $1$,
whereas $\xi_1$ and $\xi_2$ are for generic
kinematical variables for the first and second scattering, respectively.

The effective cross section $\sigma_{eff}$ can be loosely interpreted 
as a measure of the transverse correlation of the two partons inside the hadrons. 
In this paper we stick to the widely used value  $\sigma_{eff}$ = 15 mb.

We also have to use an ansatz for DPDFs, which is, 
for collinear-factorization,
\beq
D_{1, 2}(x_1,x_2,\mu) = f_1(x_1,\mu)\, f_2(x_2,\mu) \, \theta(1-x_1-x_2) \, , 
\eeq
where $D_{1, 2}(x_1,x_2,\mu)$ is the DPDF and
$f_i(x_i,\mu)$ are the ordinary PDFs. 
The subscripts $1$ and $2$ distinguish the two generic partons in the same proton.
Of course this ansatz can be automatically generalised to the case when parton
transverse momenta are included by simply including the dependence
on the transverse momentum.

Coming to DPS contributions, in principle we must include all the possible $45$ channels which
can be obtained by coupling in all possible distinct ways the $8$ channels for the $2\rightarrow 2$ SPS process, i.e.
\bea
\#1 &=& g g \rightarrow g g  \, , \quad \#5 = q \bar{q} \rightarrow q'\bar{q}'  \nn \; , \\
\#2 &=& g g \rightarrow q \bar{q} \, , \quad \#6 = q \bar{q} \rightarrow g g  \nn \;, \\
\#3 &=& q g \rightarrow q g    \, , \quad \#7 = q q \rightarrow q q \nn \; , \\
\#4 &=& q \bar{q} \rightarrow q \bar{q}   \, , \quad \#8 = q q'\rightarrow q q'   \nn \; .
\eea
We find that the pairs $(1,1)$, $(1,2)$, $(1,3)$, $(1,7)$, $(1,8)$, $(3,3)$ $(3,7)$, $(3,8)$ 
together account for more than $95$ \% of the total cross section. 
This was tested for all the sets of cuts considered in this paper.

%%%%%%%%%%%%%%%%%%%%%%%%%%%%%%%%%%%%%%%%%%%%%%%%%%%
\subsection{Comparison to the collinear approach and to ATLAS data with hard central kinematic cuts}
%%%%%%%%%%%%%%%%%%%%%%%%%%%%%%%%%%%%%%%%%%%%%%%%%%%

Our HEF calculation was first tested against the 8 TeV
data recently reported by the ATLAS collaboration \cite{Aad:2015nda}. 
The kinematic cuts are $p_T > 100 $ GeV for the leading jet and $p_T > 64 $ GeV for the first three subleading jets; in addition
$|\eta| < 2.8$ is the pseudorapidity cut and $\Delta R > 0.65$ is the constraint on the jet cone radius parameter.

We employ the running NLO $\alpha_s$ coming with the MSTW2008 sets.
For such hard central cuts, not much difference is expected between the two approaches
and indeed we find none. Also, DPS effects are irrelevant with this kinematics
and this is confirmed by our analysis, which is presented in much more detail in \cite{Kutak:2016mik}

The collinear factorization performs slightly better for intermediate values and HEF
does a better job for the last bins, except for the $4$th jet.
All in all, both approaches are consistent with the data in this kinematic region.

%%%%%%%%%%%%%%%%%%%%%%%%%%%%%%%%%%%%%%%%%%%%%%%%%%%%%%%%%%%%%%%%%%%%%%%%
\subsection{Comparison to CMS data with softer cuts}
%%%%%%%%%%%%%%%%%%%%%%%%%%%%%%%%%%%%%%%%%%%%%%%%%%%%%%%%%%%%%%%%%%%%%%%%

As discussed in Ref.~\cite{Maciula:2015vza}, so far the only experimental analysis 
of four-jet production relevant for the DPS studies was realized
by the CMS collaboration \cite{Chatrchyan:2013qza}. 
The cuts used in this analysis are $p_T> 50$ GeV for the first and
second jets, $p_T > 20$ GeV for the third and fourth jets,
$|\eta| < 4.7$ and the jet cone radius parameter $\Delta R > 0.5$. 
In the rest of this section, we present our results for such cuts.

As for the total cross section for the four jet production, the experimental and theoretical LO results are:
\bea
\text{CMS collaboration} : &&
\sigma_{tot} = 330 \pm 5\, (\text{stat.}) \pm 45\, (\text{syst.})\,  nb
\nn \\
\text{LO collinear factorization}: &&
\sigma_{SPS} = 697\,  nb\, , \quad \sigma_{DPS} = 125\, nb \, ,  \quad \sigma_{tot} = 822\, nb 
\nn \\
\text{LO HEF $k_{T}$-factorization}: &&
\sigma_{SPS} = 548\,  nb\, , \quad \sigma_{DPS} = 33\, nb \, , \quad \sigma_{tot} = 581\, nb
\label{sigma_CMS}
\eea
It is apparent that the LO results need refinements from NLO contributions,
much more than it does in the case of the ATLAS hard cuts, 
as we are of course not that deep into the perturbative region.
For this reason, in the following we will always perform comparisons only to 
data normalised to the total (SPS+DPS) cross sections.
we find that this is better than introducing fixed K-factors, whose
phase-space dependence is never really under control. 
What is immediately apparent in the HEF total cross section 
is the dramatic damping of the DPS contribution with respect to the collinear case. 
The effect of the damping is of kinematical nature and can be understood
by an analogy of a similar effect first observed in NLO
jet photoproduction at HERA \cite{Frixione:1997ks}.
The point is that the emission of gluon radiation, which is taken into account
via the TMDs in our approach and via the real contribution in a collinear NLO calculation, 
alters the exact momentum balance of the final state two-jet system, so that a lot of events are not 
taken into account for the higher transverse momentum just above the cut. 
%
%--------------------------------------------------------------------------------
\begin{figure}[h]
\begin{center}
\begin{minipage}{0.47\textwidth}
 \centerline{\includegraphics[width=1.0\textwidth]{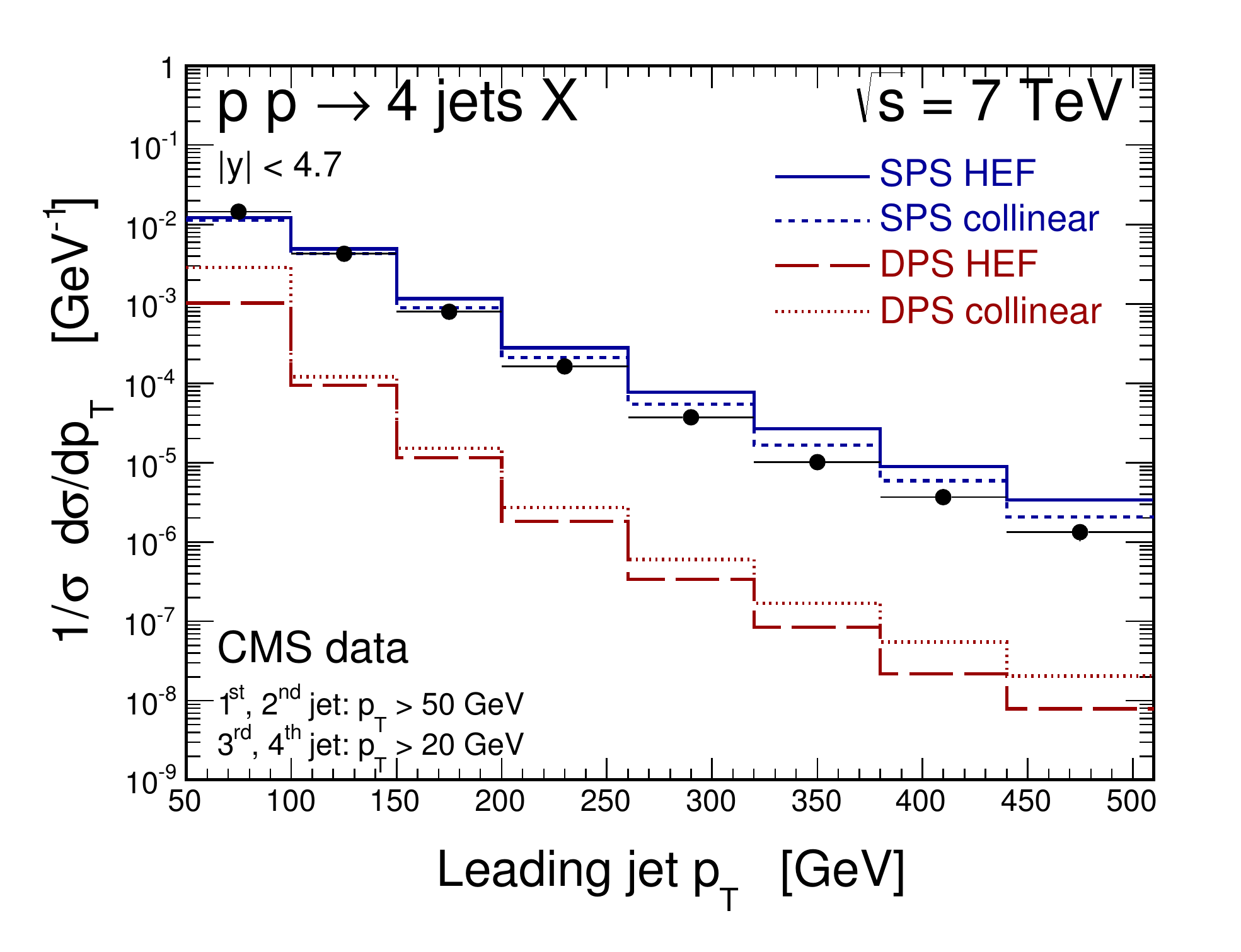}}
\end{minipage}
\hspace{0.5cm}
\begin{minipage}{0.47\textwidth}
 \centerline{\includegraphics[width=1.0\textwidth]{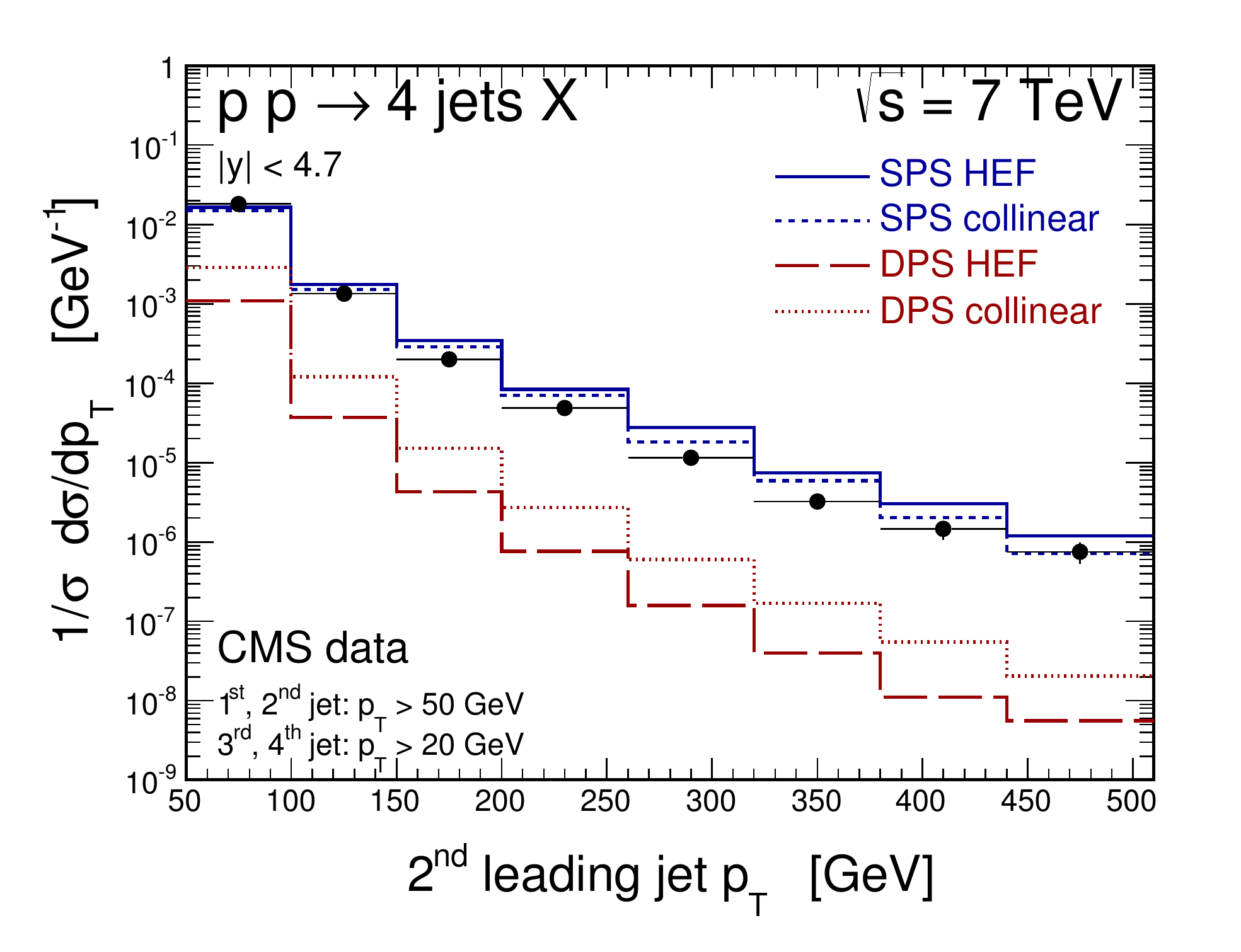}}
\end{minipage}
\end{center}
\caption{
Comparison of the LO collinear and HEF predictions to
the CMS data for the 1st and 2nd leading jets.}
\label{CMS_pT_12}
\end{figure}
%---------------------------------------------------------------------------------
%
%
In Fig.~\ref{CMS_pT_12} we compare the predictions
in HEF to the CMS data for the 1st and 2nd leading jets transverse momenta spectra. 
Here both the SPS and DPS contributions are normalized to the total cross section,
i.e. the sum of the SPS and DPS contributions.
In all cases the renormalized transverse momentum distributions agree quite well 
with the CMS data. 

%%%%%%%%%%%%%%%%%%%%%%%%%%%%%%%%%%%
\subsection{HEF predictions for a possible set of asymmetric cuts }
%%%%%%%%%%%%%%%%%%%%%%%%%%%%%%%%%%%

Moving from the previous considerations, we present our results for four-jet
production by employing also another set of cuts, which are asymmetric with respect to the
final state transverse momenta.
Specifically, we require $p_T > 35$ GeV for the leading jet, 
$p_T > 20$ GeV for all the other jets and we stick to $|\eta| < 4.7$, $\Delta R > 0.65$
for rapidity and jet size parameter.
An experimental analysis with such cuts is not available: while the CMS collaboration did
perform the analysis for soft enough cuts as to allow for significant DPS contributions
to show up, they did not impose such asymmetry, as discussed above, whereas
both analysis presented by the ATLAS collaboration employs too hard
cuts for multi-parton interactions to be any significant at all \cite{Aad:2015nda}.
Of course it would be desirable to have such an analysis  in the future.

The theoretical total cross sections for these cuts for four-jet production are:
\bea
\text{LO collinear factorization}: &&
\sigma_{SPS} = 1969 \,  nb\, , \quad \sigma_{DPS} = 514\, nb 
\nn \\
\text{LO HEF $k_{T}$-factorization}: &&
\sigma_{SPS} = 1506 \,  nb\, , \quad \sigma_{DPS} =  297 \, nb
\eea
When comparing to (\ref{sigma_CMS}), it is apparent that now the drop in the total cross section for DPS 
when moving from LO collinear to HEF approach is considerably smaller, as argued.

%
%---------------------------------------------------------------------------------
\begin{figure}[h]
\begin{center}
\begin{minipage}{0.47\textwidth}
 \centerline{\includegraphics[width=1.0\textwidth]{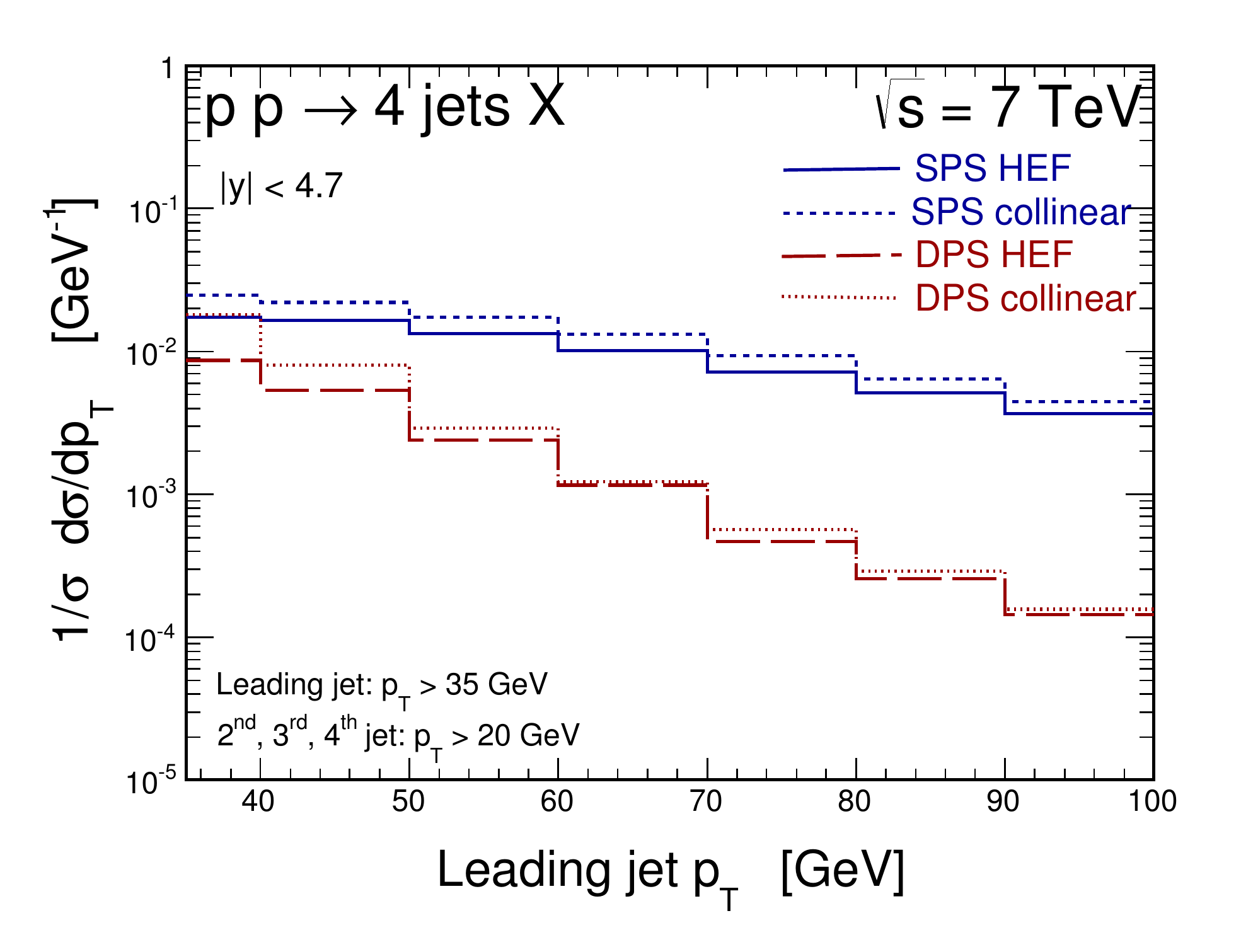}}
\end{minipage}
\hspace{0.5cm}
\begin{minipage}{0.47\textwidth}
 \centerline{\includegraphics[width=1.0\textwidth]{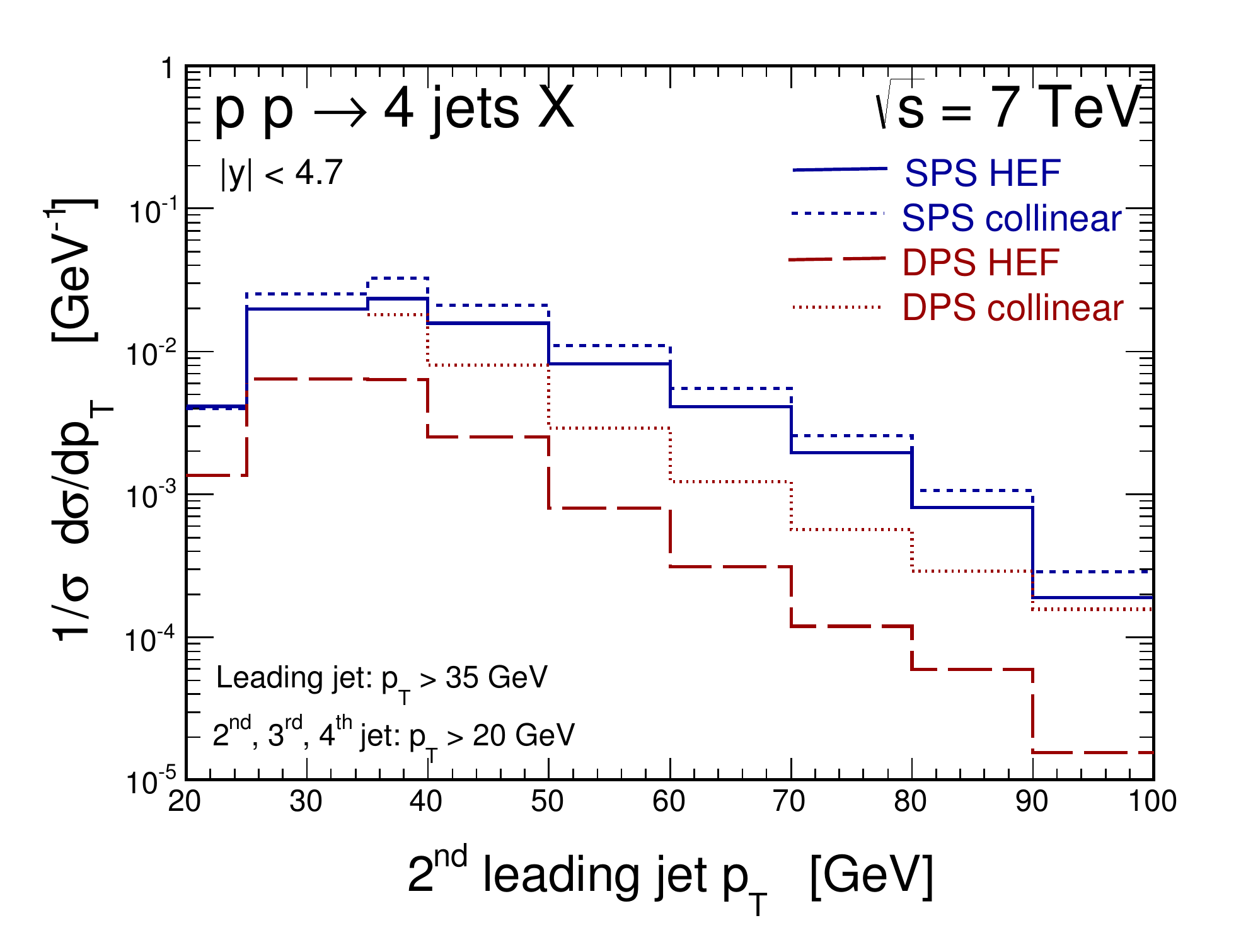}}
\end{minipage}
\end{center}
\caption{
LO collinear and HEF predictions for the 1st and 2nd 
leading jets with the asymmetric cuts.}
\label{Asymm_pT_12}
\end{figure}
%----------------------------------------------------------------------------------

In Fig.~\ref{Asymm_pT_12} we show our predictions
for the normalized transverse momentum distributions with the new set of cuts.

%%%%%%%%%%%%%%%%%%%%%%%%%%%%%%
\section{Conclusions}%%
%%%%%%%%%%%%%%%%%%%%%%%%%%%%%%

In the present work we have compared the perturbative predictions 
for four-jet production at the LHC in leading-order collinear and high-energy ($k_T$-)factorization.
While we find that there is no significant difference between the collinear
and HEF approach for hard central cuts, significant differences show up,
especially for DPS, when the cuts on the transverse momenta are lowered.
We agree with with Ref.~\cite{Maciula:2015vza} that lowering the cut 
in transverse momenta can significantly enhance the experimental sensitivity to DPS but 
we also observe that HEF severely tames this effect for symmetric cuts,
due to gluon-emission effects which alter the transverse-momentum balance between final state partons. 
We have found that the damping is not present when cuts are not identical.
A more complete treatment of the subjects addressed in this proceeding can be found in \cite{Kutak:2016mik}. 

%%%%%%%%%%%%%%%%%%%%%%%%%%%%%%%%%%%%%%%%%%%%%%%%%%%%%%%%%%%%%%%%%
\section*{Acknowledgments}
The work of M.S. has been supported by Narodowe Centrum Nauki
with Sonata Bis grant DEC-2013/10/E/ST2/00656.
M.S. also thanks the "Angelo della Riccia" foundation for support.
%%%%%%%%%%%%%%%%%%%%%%%%%%%%%%%%%%%%%%%%%%%%%%%%%%%%%%%%%%%%%%%%%


\begin{thebibliography}{10}
\newcommand{\enquote}[1]{``#1''}
\providecommand{\url}[1]{\texttt{#1}}
\providecommand{\urlprefix}{URL }
\providecommand{\eprint}[2][]{\url{#2}}

\bibitem{Maciula:2015vza}
Rafal Maciula and Antoni Szczurek, \enquote{{Searching for and exploring
  double-parton scattering effects in four-jet production at the LHC},}
  \emph{Phys. Lett.}, \textbf{B749}(2015), 57, \eprint{1503.08022}.

\bibitem{Kimber:2001sc}
M.~A. Kimber, Alan~D. Martin, and M.~G. Ryskin, \enquote{{Unintegrated parton
  distributions},} \emph{Phys. Rev.}, \textbf{D63}(2001), 114027,
  \eprint{hep-ph/0101348}.

\bibitem{Kimber:1999xc}
M.~A. Kimber, Alan~D. Martin, and M.~G. Ryskin, \enquote{{Unintegrated parton
  distributions and prompt photon hadroproduction},} \emph{Eur. Phys. J.},
  \textbf{C12}(2000), 655, \eprint{hep-ph/9911379}.

\bibitem{vanHameren:2014iua}
A.~van Hameren, \enquote{{BCFW recursion for off-shell gluons},} \emph{JHEP},
  \textbf{07}(2014), 138, \eprint{1404.7818}.

\bibitem{Bury:2015dla}
M.~Bury and A.~van Hameren, \enquote{{Numerical evaluation of multi-gluon
  amplitudes for High Energy Factorization},} \emph{Comput. Phys. Commun.},
  \textbf{196}(2015), 592, \eprint{1503.08612}.

\bibitem{vanHameren:2015bba}
A.~van Hameren and M.~Serino, \enquote{{BCFW recursion for TMD parton
  scattering},} \emph{JHEP}, \textbf{07}(2015), 010, \eprint{1504.00315}.

\bibitem{Aad:2015nda}
Georges Aad \emph{et~al.} (ATLAS), \enquote{{Measurement of four-jet
  differential cross sections in $\sqrt{s}=8$ TeV proton-proton collisions
  using the ATLAS detector},} \emph{JHEP}, \textbf{12}(2015), 105,
  \eprint{1509.07335}.

\bibitem{Kutak:2016mik}
Krzysztof Kutak, Rafal Maciula, Mirko Serino, Antoni Szczurek, and Andreas van
  Hameren, \enquote{{Four-jet production in single- and double-parton
  scattering within high-energy factorization},} \emph{JHEP},
  \textbf{04}(2016), 175, \eprint{1602.06814}.

\bibitem{Chatrchyan:2013qza}
Serguei Chatrchyan \emph{et~al.} (CMS), \enquote{{Measurement of four-jet
  production in proton-proton collisions at $\sqrt{s}=7$??TeV},}
  \emph{Phys. Rev.}, \textbf{D89}(2014)~(9), 092010, \eprint{1312.6440}.

\bibitem{Frixione:1997ks}
Stefano Frixione and Giovanni Ridolfi, \enquote{{Jet photoproduction at HERA},}
  \emph{Nucl. Phys.}, \textbf{B507}(1997), 315, \eprint{hep-ph/9707345}.

\end{thebibliography}
\end{document}